\documentclass[prl,superscriptaddress,twocolumn,preprintnumbers,nofootinbib,%
  amsmath,amssymb,longbibliography]{revtex4-1}
  
\usepackage[breaklinks,colorlinks=true]{hyperref}
\usepackage[T1]{fontenc}
\usepackage[utf8]{inputenc}
\usepackage{mathtools}
\usepackage{amsmath,amsthm,amssymb}
\usepackage{amsmath,bm}
\usepackage{bbold} 
\usepackage{dsfont}
\usepackage{lipsum}
\usepackage{color}
\usepackage{hyperref}
\usepackage{bm}
\usepackage{xspace}
\usepackage{makecell}
\usepackage{bibunits}
\defaultbibliography{main}
\defaultbibliographystyle{apsrev4-1}
\allowdisplaybreaks

\usepackage[dvipsnames]{xcolor}
\newcommand\myshade{85}
\colorlet{mylinkcolor}{BrickRed}
\colorlet{mycitecolor}{NavyBlue}
\colorlet{myurlcolor}{Aquamarine}
\hypersetup{
  linkcolor  = mylinkcolor!\myshade!black,
  citecolor  = mycitecolor!\myshade!black,
  urlcolor   = myurlcolor!\myshade!black,
  colorlinks = true,
}

\def\compiletikz{0}

\if1\compiletikz
\usepackage{tikz}
\usepackage{pgfplots}
\pgfplotsset{compat = newest}
\usepackage{pgfplotstable}
\usepackage{circuitikz}

\usetikzlibrary{positioning, calc, matrix, fit, arrows, arrows.meta, arrows.new, shapes.misc, shapes.geometric, backgrounds}
\usetikzlibrary{pgfplots.external}
\usetikzlibrary{colorbrewer}
\usetikzlibrary{external}
\usepgfplotslibrary{colorbrewer}
\usepgfplotslibrary{groupplots}
\usepgfplotslibrary{statistics}
\usepgfplotslibrary{fillbetween}
\usepgfplotslibrary{colormaps}

\usepackage{filemod}

\pgfplotsset{grid style={dashed,gray!40}}

\pgfplotsset{
  /pgfplots/scatter legend/.style={
    /pgfplots/legend image code/.code={\draw[##1,mark size=0.9pt,yshift=-0.1em, xshift=7pt] plot coordinates {
      (-0.1em, 0.1em)
      (0.2em, 0.45em)
      (0.3em, -0.05em)
      (0.6em, 0.3em)
    };},
  },
  /pgfplots/line and fill/.style={%
    legend image code/.code={%
      \fill [
        ##1,
        opacity=0.3,
        draw=none,
        rounded corners=0pt,
      ] (0mm,-1.2mm) rectangle (3mm,1.2mm);
      \draw [##1,fill=none, thick] (0mm,0mm) -- (3mm,0mm);
    },
  },
}

\pgfplotsset{
    jitter/.style={
      x filter/.code={\pgfmathparse{\pgfmathresult+rnd*#1-0.5*#1}}
    },
    jitter/.default=0.2,
    scatterstyle/.style={
      only marks,
      jitter, 
      mark size=0.9pt,
    },
    boxstyle/.style={
      boxplot,
      boxplot/box extend=0.7,
      boxplot/whisker range=3,
      fill,
      fill opacity=0.4,
      line width=0.7pt,
    },
}

\definecolor{BrewerRed}{RGB}{228,26,28}
\definecolor{BrewerGreen}{RGB}{77,175,74}
\definecolor{BrewerBlue}{RGB}{55,126,184}
\definecolor{BrewerPurple}{RGB}{152,78,163}
\definecolor{BrewerOrange}{RGB}{255,127,0}
\definecolor{BrewerYellow}{RGB}{215,215,41}
\definecolor{BrewerGray}{RGB}{150,150,150}

\pgfplotsset{
    /pgfplots/layers/axlayers/.define layer set={
        axis background,axis grid,pre main,main,axis ticks,axis lines,axis tick labels,
        axis descriptions,axis foreground
    }{/pgfplots/layers/standard},
}

\newcommand\lineboxplot[6]{%
  \pgfplotstableread[col sep=comma]{#1}\data
  \addplot[boxstyle, #5] table [x expr=1, y=#3, col sep=comma] {\data};
  \addplot[boxstyle, #6] table [x expr=2, y=#4, col sep=comma] {\data};
  \pgfplotsforeachungrouped \i in {0,...,#2}{
    \pgfplotstablegetelem{\i}{#4}\of\data
    \pgfmathsetmacro{\lsd}{\pgfplotsretval}
    \pgfplotstablegetelem{\i}{#3}\of\data
    \pgfmathsetmacro{\pla}{\pgfplotsretval}
    \edef\donode{\noexpand\node[mydot, #5, fill=#5] (pla\i) at (axis cs: 1+0.1*rand, \pla) {};}
    \donode
    \edef\donode{\noexpand\node[mydot, #6, fill=#6] (lsd\i) at (axis cs: 2+0.1*rand, \lsd) {};}
    \donode
  }
  \pgfplotsforeachungrouped \i in {0,...,#2}{
    \edef\doplot{\noexpand\draw[linestyle] (pla\i) -- (lsd\i);}
    \doplot
  }
}

\newcommand{\errorbandlegend}[8][]{%
\pgfplotstableread[col sep=comma]{#2}\datatable
  \addplot [name path=pluserror#4,draw=none,no markers,forget plot]
    table [x expr=\thisrow{#3},y expr=\thisrow{#4}+\thisrow{#5}] {\datatable};
  \addplot [name path=minuserror#4,draw=none,no markers,forget plot]
    table [x expr=\thisrow{#3},y expr=\thisrow{#4}-\thisrow{#5}] {\datatable};
  \addplot [forget plot,fill=#6,opacity=#7]
    fill between[on layer={},of=pluserror#4 and minuserror#4];
  \addplot [#8,thick,no markers]
    table [x expr=\thisrow{#3},y={#4}] {\datatable};
  \addlegendentry{#1}
}

\colorlet{ErasureColourLight}{Paired-A}
\colorlet{ErasureColour}{Paired-B}
\colorlet{TransferColourLight}{Paired-E}
\colorlet{TransferColour}{Paired-F}
\colorlet{HighOrderColourLight}{Paired-C}
\colorlet{HighOrderColour}{Paired-D}
\colorlet{ReversibleColour}{gray}

\fi

\newcommand\subref[2]{\hyperref[#1]{\ref*{#1}#2}}

\newcommand\PI[2]{\ensuremath{I_{\partial}^{#1\rightarrow #2}}\xspace}

\newcommand{\phiid}{\ensuremath{\Phi\mathrm{ID}}\xspace}

\newcommand\rtr{\ensuremath{\texttt{Red}\!\rightarrow\!\texttt{Red}}\xspace}
\newcommand\rtx{\ensuremath{\texttt{Red}\!\rightarrow\!\texttt{Un\textsuperscript{X}}}\xspace}
\newcommand\rty{\ensuremath{\texttt{Red}\!\rightarrow\!\texttt{Un\textsuperscript{Y}}}\xspace}
\newcommand\rts{\ensuremath{\texttt{Red}\!\rightarrow\!\texttt{Syn}}\xspace}
\newcommand\xtr{\ensuremath{\texttt{Un\textsuperscript{X}}\!\rightarrow\!\texttt{Red}}\xspace}
\newcommand\xtx{\ensuremath{\texttt{Un\textsuperscript{X}}\!\rightarrow\!\texttt{Un\textsuperscript{X}}}\xspace}
\newcommand\xty{\ensuremath{\texttt{Un\textsuperscript{X}}\!\rightarrow\!\texttt{Un\textsuperscript{Y}}}\xspace}
\newcommand\xts{\ensuremath{\texttt{Un\textsuperscript{X}}\!\rightarrow\!\texttt{Syn}}\xspace}
\newcommand\ytr{\ensuremath{\texttt{Un\textsuperscript{Y}}\!\rightarrow\!\texttt{Red}}\xspace}
\newcommand\ytx{\ensuremath{\texttt{Un\textsuperscript{Y}}\!\rightarrow\!\texttt{Un\textsuperscript{X}}}\xspace}
\newcommand\yty{\ensuremath{\texttt{Un\textsuperscript{Y}}\!\rightarrow\!\texttt{Un\textsuperscript{Y}}}\xspace}
\newcommand\yts{\ensuremath{\texttt{Un\textsuperscript{Y}}\!\rightarrow\!\texttt{Syn}}\xspace}
\newcommand\str{\ensuremath{\texttt{Syn}\!\rightarrow\!\texttt{Red}}\xspace}
\newcommand\stx{\ensuremath{\texttt{Syn}\!\rightarrow\!\texttt{Un\textsuperscript{X}}}\xspace}
\newcommand\sty{\ensuremath{\texttt{Syn}\!\rightarrow\!\texttt{Un\textsuperscript{Y}}}\xspace}
\newcommand\sts{\ensuremath{\texttt{Syn}\!\rightarrow\!\texttt{Syn}}\xspace}

\begin{document}

\title{Information decomposition reveals hidden high-order contributions\\ to temporal irreversibility}

\author{Andrea~I~Luppi}
\email{al857@cam.ac.uk}
\affiliation{University Division of Anaesthesia and Department of Clinical Neurosciences, University of Cambridge, Cambridge, UK}
\affiliation{Montreal Neurological Institute, McGill University, Montreal, Canada}

\author{Fernando E. Rosas}
\email{f.rosas@sussex.ac.uk}
\affiliation{Department of Informatics, University of Sussex, Brighton, UK}
\affiliation{Centre for Psychedelic Research, Department of Brain Science, and Centre for Complexity Science, Imperial College London, London, UK}
\affiliation{Centre for Eudaimonia and Human Flourishing, Linacre College, University of Oxford, Oxford, UK}

\author{Gustavo Deco}
\affiliation{Center for Brain and Cognition, Department of Information and Communication Technologies, Universitat Pompeu Fabra, Barcelona, Spain}
\affiliation{Institució Catalana de la Recerca i Estudis Avançats (ICREA), Barcelona, Spain}

\author{Morten L. Kringelbach}
\affiliation{Centre for Eudaimonia and Human Flourishing, Linacre College, University of Oxford, Oxford, UK}
\affiliation{Department of Psychiatry, University of Oxford, Oxford, UK}
\affiliation{Center for Music in the Brain, Department of Clinical Medicine, Aarhus University, Aarhus, Denmark}

\author{Pedro A.M. Mediano}
\email{p.mediano@imperial.ac.uk}
\affiliation{Department of Computing, Imperial College London, London, UK}

\newtheorem{definition}{Definition}
\newtheorem{theorem}{Theorem}
\newtheorem{lemma}{Lemma}
\newtheorem{proposition}{Proposition}
\newtheorem{corollary}{Corollary}
\newtheorem{example}{Example}
\newtheorem{remark}{Remark}

\begin{abstract}

\noindent

\noindent
Temporal irreversibility, often referred to as the `arrow of time,' is a fundamental concept in statistical mechanics. Markers of irreversibility also provide a powerful characterisation of information processing in biological systems.  However, current approaches tend to describe temporal irreversibility in terms of a single scalar quantity, without disentangling the underlying dynamics that contribute to irreversibility. Here we propose a broadly applicable information-theoretic framework to characterise the arrow of time in multivariate time series, which yields qualitatively different types of irreversible information dynamics. This multidimensional characterisation reveals previously unreported high-order modes of irreversibility, and establishes a formal connection between recent heuristic markers of temporal irreversibility and metrics of information processing. We demonstrate the prevalence of high-order irreversibility in the hyperactive regime of a biophysical model of brain dynamics, showing that our framework is both theoretically principled and empirically useful.
This work challenges the view of the arrow of time as a monolithic entity, enhancing both our theoretical understanding of irreversibility and our ability to detect it in practical applications. 
\end{abstract}

\maketitle

\begin{bibunit}

The stream of time seems to steadily flow forward and never go back, creating an asymmetry between past and future often referred to as the `arrow of time'~\cite{schulman1997time}. 
While temporal irreversibility is typically studied in thermodynamics via entropy production~\cite{esposito2010three}, a promising complementary line of work investigates irreversibility in terms of information processing. 
For example, physicists have discovered that some types of information-processing operations can be implemented reversibly (e.g. the Toffoli logic gate~\cite{toffoli1980reversible}), while others are intrinsically irreversible (e.g. information erasure~\cite{landauer1961irreversibility,bennett1982thermodynamics}). 
Clarifying the relationship between information processing and irreversibility has led to important advances in the thermodynamics of information~\cite{parrondo2015thermodynamics}, enabling for example the design of novel engines that can transform information into work~\cite{boyd2016identifying}. 
These breakthroughs have also inspired work in other fields, including neuroscience, showing that statistical metrics of irreversibility applied to neural recordings can provide insights about cognition and pathology across species \cite{deco2022insideout, kringelbach2023toward, tewarie2023nonreversibility, cruzat2023temporal, deco2022arrow, guzman2023doc, bernardi2023ocd, deLaFuente2023cercor}. However, despite the effectiveness of these markers, their heuristic nature makes them unable to provide insight about which properties of the system's dynamics determine its irreversibility --- or lack thereof.

Here we introduce a novel and broadly applicable information-theoretic framework to quantify irreversibility based on high-order statistics, which is particularly well-suited to the practical analysis of multivariate time series data. While most approaches to study temporal irreversibility describe it as a scalar quantity~\cite{zanin2021algorithmic}, our framework provides a multidimensional characterisation that identifies \emph{qualitatively different types of information phenomena that contribute to irreversibility} in multivariate processes. 

The proposed framework places existing heuristic markers of irreversibility on the firm theoretical footing of information dynamics, while establishing a common information-theoretic language to relate them to other metrics of information processing and dynamical complexity. 
Furthermore, the framework reveals the existence of high-order modes of irreversibility not previously reported, and shows that they dominate the irreversibility observed beyond the critical point of a well-known biophysical model of brain dynamics.

\begin{figure}[t]
\centering

\if1\compiletikz
  \tikzsetnextfilename{ExampleSystems}%
  \filemodCmp{tikz/ExampleSystems.tex}{tikz/external/ExampleSystems.pdf}%
    {\tikzset{external/remake next}}{}%
  \begin{tikzpicture}[tight background]

\tikzstyle{var} = [circle,fill=BrewerBlue!20,draw=black, inner sep=1pt,
                   minimum size=13pt, node distance=1]

\tikzstyle{sys} = [inner sep=0pt]

\tikzstyle{lbl} = [above, font=\small]

\definecolor{BrewerRed}{RGB}{228,26,28}
\definecolor{BrewerGreen}{RGB}{77,175,74}
\definecolor{BrewerBlue}{RGB}{55,126,184}
\definecolor{BrewerPurple}{RGB}{152,78,163}
\definecolor{BrewerOrange}{RGB}{255,127,0}

\def\hsep{1.7}
\def\ysep{0.7}

\tikzstyle{xor} = [circle,fill=black,draw=black, inner sep=0pt,
minimum size=4pt, anchor=center]

\tikzstyle{arr} = [-latex]
\tikzstyle{doublearr} = [{Classical TikZ Rightarrow[length=4pt, width=4.5pt]}-{Classical TikZ Rightarrow[length=4pt, width=5pt]}, double]

\tikzset{
  system/.pic={
      \node[var, fill=#1] at (    0,    0) (n1) {};
      \node[var, fill=#1] at (\hsep,    0) (m1) {};
      \node[var, fill=#1] at (    0,\ysep) (n2) {};
      \node[var, fill=#1] at (\hsep,\ysep) (m2) {};
      \coordinate (bot) at ($(n1)!0.5!(m1)$);
      \coordinate (lat) at ($(n1)!0.5!(n2)$);
      \coordinate (mid) at (bot |- lat);
  },
}


\def\rowsep{2.6cm}
\def\colsep{3.3cm}

\node[sys, anchor=center, label={[lbl]Erasure}] (erasure) {
  \begin{tikzpicture}
  \path (0,0) pic {system=ErasureColourLight};
  \node[xor] (x) at (mid) {};
  \draw (n1) -- (x);
  \draw (n2) -- (x);
  \path[arr] (x) edge (m1);
  \end{tikzpicture}
}; 

\node[sys, anchor=west, right=\colsep of erasure, label={[lbl]Copy}] (copy) at (0,0) {
  \begin{tikzpicture}
  \path (0,0) pic {system=ErasureColourLight};
  \node[xor] (x) at (mid) {};
  \draw (n1) -- (x);
  \path[arr] (x) edge (m1) (x) edge (m2);
  \end{tikzpicture}
}; 

\coordinate (l1) at ($(erasure) - (0, \rowsep)$);
\coordinate (l2) at ($(l1) - (0, \rowsep)$);

\coordinate (m) at ($(erasure.east)!0.5!(copy.west)$);
\node[anchor=south, outer sep=5pt] (time) at (m) {
  \begin{tikzpicture}
    \node[inner sep=0pt, outer sep=2pt] (txt) {\footnotesize\textcolor{red}{\makecell[c]{Time\\reversal}}};
    \node[anchor=west, inner sep=0pt, outer sep=2pt] (img) at (txt.east) {
  \begin{tikzpicture}
  \newlength\R
  \setlength\R{0.3cm}
  \draw[line width=0.4pt] (0,0) circle (\R);
	\foreach \i in {1,2,...,60}{
		\def\angle{\i*6}
    \draw[line width=0.3pt] (\angle:\R) -- (\angle:0.9\R);
	}
%
	\foreach \i in {1,2,...,12}{
		\def\angle{\i*-30+90}
    \draw[line width=0.3pt] (\angle:\R) -- (\angle:0.8\R);
	};
%
  \draw[line width=0.7pt] (0,0) -- (80:0.4\R);
%
  \draw[line width=0.5pt] (0,0) -- (150:0.6\R);

  \draw[-{Latex[length=1.9pt, width=1.9pt]}, red] (80:1.2\R) arc (80:150:1.2\R);
  \begin{pgfonlayer}{background}
    \filldraw[fill=red!30!white, draw=none] (0,0) -- (80:\R) arc (80:150:\R) -- cycle;
  \end{pgfonlayer}
%
  \end{tikzpicture}
};
\end{tikzpicture}
}; 

\node[sys, anchor=center, label={[lbl]$X\!\rightarrow\!Y$ transfer}] at (l1 -| erasure.center) (xytransfer) {
  \begin{tikzpicture}
  \path (0,0) pic {system=TransferColourLight};
  \path[arr] (n2) edge (m1);
  \end{tikzpicture}
}; 

\node[sys, anchor=west, label={[lbl]$Y\!\rightarrow\!X$ transfer}] at (xytransfer -| copy.west) (yxtransfer) {
  \begin{tikzpicture}
  \path (0,0) pic {system=TransferColourLight};
  \path[arr] (n1) edge (m2);
  \end{tikzpicture}
}; 

\node[sys, anchor=center, label={[lbl]High- to low-order}] at (l2 -| erasure.center) (downward) {
  \begin{tikzpicture}
  \path (0,0) pic {system=HighOrderColourLight};
  \node[fit=(n1)(n2), rounded corners=9pt, inner sep=4pt, draw, dashed] (l) {};
  \node[fit=(m1)(m2), rounded corners=9pt, inner sep=4pt, draw=none, dashed] {};
  \node[xor] (x) at (mid) {};
  \path[arr] (x) edge (m1) (x) edge (m2);
  \draw (l) -- (x);
  \end{tikzpicture}
}; 

\node[sys, anchor=center, label={[lbl]Low- to high-order}] at (downward -| copy.center) (upward) {
  \begin{tikzpicture}
  \path (0,0) pic {system=HighOrderColourLight};
  \node[fit=(m1)(m2), rounded corners=9pt, inner sep=4pt, draw, dashed] (l) {};
  \node[fit=(n1)(n2), rounded corners=9pt, inner sep=4pt, draw=none, dashed] {};
  \node[xor] (x) at (mid) {};
  \draw (n1) -- (x);
  \draw (n2) -- (x);
  \path[arr] (x) edge (l);
  \end{tikzpicture}
}; 


\tikzstyle{arr} = [{Implies}-{Implies}, double]
\draw[doublearr] (erasure -| downward.east) -- (copy -| upward.west);
\draw[doublearr] (xytransfer -| downward.east) -- (yxtransfer -| upward.west);
\draw[doublearr] (downward -| downward.east) -- (upward -| upward.west);

\node[inner sep=0pt, anchor=center, fit=(erasure)(time)(upward)] (bbox) {};

\node[below=1cm of bbox.south] (matrix) {
\begin{tikzpicture}
\newlength\matsize
\setlength\matsize{3.5cm}
\node[anchor=east] at (0,0.875*\matsize) (ps) {\texttt{Syn}};
\node[anchor=east] at (0,0.625*\matsize) (py) {\texttt{Un\textsuperscript{Y}}};
\node[anchor=east] at (0,0.375*\matsize) (px) {\texttt{Un\textsuperscript{X}}};
\node[anchor=east] at (0,0.125*\matsize) (pr) {\texttt{Red}};
\node[inner sep=0pt, outer sep=3pt, fit=(ps)(px)(py)(pr)] (p) {};
\node[anchor=south, rotate=90] at (p.west) {\scshape\bfseries PAST};
\node[anchor=north east, rotate=45] at (0.125*\matsize, 0) (fs) {\texttt{Red}};
\node[anchor=north east, rotate=45] at (0.375*\matsize, 0) (fy) {\texttt{Un\textsuperscript{X}}};
\node[anchor=north east, rotate=45] at (0.625*\matsize, 0) (fx) {\texttt{Un\textsuperscript{Y}}};
\node[anchor=north east, rotate=45] at (0.875*\matsize, 0) (fr) {\texttt{Syn}};
\node[inner sep=0pt, outer sep=3pt, fit=(fs)(fx)(fy)(fr)] (f) {};
\node[anchor=north, rotate=0] at (f.south) {\scshape\bfseries FUTURE};
%
\node[minimum size=0.25*\matsize, draw=none, fill=HighOrderColourLight, anchor=center] at (0.875*\matsize, 0.125\matsize) (str) {};
\node[minimum size=0.25*\matsize, draw=none, fill=HighOrderColourLight, anchor=center] at (0.875*\matsize, 0.375\matsize) (stx) {};
\node[minimum size=0.25*\matsize, draw=none, fill=HighOrderColourLight, anchor=center] at (0.875*\matsize, 0.625\matsize) (sty) {};
\node[minimum size=0.25*\matsize, draw=none, fill=HighOrderColourLight, anchor=center] at (0.125*\matsize, 0.875\matsize) (rts) {};
\node[minimum size=0.25*\matsize, draw=none, fill=HighOrderColourLight, anchor=center] at (0.375*\matsize, 0.875\matsize) (xts) {};
\node[minimum size=0.25*\matsize, draw=none, fill=HighOrderColourLight, anchor=center] at (0.625*\matsize, 0.875\matsize) (yts) {};
\node[minimum size=0.25*\matsize, draw=none, fill=TransferColourLight, anchor=center] at (0.625*\matsize, 0.375\matsize) (ytx) {};
\node[minimum size=0.25*\matsize, draw=none, fill=TransferColourLight, anchor=center] at (0.375*\matsize, 0.625\matsize) (xty) {};
\node[minimum size=0.25*\matsize, draw=none, fill=ErasureColourLight, anchor=center] at (0.625*\matsize, 0.125\matsize) (ytr) {};
\node[minimum size=0.25*\matsize, draw=none, fill=ErasureColourLight, anchor=center] at (0.375*\matsize, 0.125\matsize) (xtr) {};
\node[minimum size=0.25*\matsize, draw=none, fill=ErasureColourLight, anchor=center] at (0.125*\matsize, 0.375\matsize) (rtx) {};
\node[minimum size=0.25*\matsize, draw=none, fill=ErasureColourLight, anchor=center] at (0.125*\matsize, 0.625\matsize) (rty) {};
\node[minimum size=0.25*\matsize, draw=none, fill=ReversibleColour!30, anchor=center] at (0.125*\matsize, 0.125\matsize) (rtr) {};
\node[minimum size=0.25*\matsize, draw=none, fill=ReversibleColour!30, anchor=center] at (0.375*\matsize, 0.375\matsize) (xtx) {};
\node[minimum size=0.25*\matsize, draw=none, fill=ReversibleColour!30, anchor=center] at (0.625*\matsize, 0.625\matsize) (yty) {};
\node[minimum size=0.25*\matsize, draw=none, fill=ReversibleColour!30, anchor=center] at (0.875*\matsize, 0.875\matsize) (sts) {};
%
\node[inner sep=0pt, outer sep=0pt, draw=black, minimum size=\matsize, anchor=south west] at (0,0) (mat) {};
\foreach \x in {0.25,0.5,0.75}{
  \draw[black, thin] (0,\x*\matsize) -- (\matsize,\x*\matsize);
  \draw[black, thin] (\x*\matsize,0) -- (\x*\matsize,\matsize);
};
%
\def\arrowsep{3pt}
\draw[doublearr] ($(rts.center)+(0,-\arrowsep)$) -- ($(str.center)+(0,-\arrowsep)$);
\draw[doublearr] (xts.center) -- (stx.center);
\draw[doublearr] (yts.center) -- (sty.center);
\draw[doublearr] ($(ytx.center)+(0,\arrowsep)$) -- ($(xty.center)+(0,\arrowsep)$);
\draw[doublearr] (rtx.center) -- (xtr.center);
\draw[doublearr] (rty.center) -- (ytr.center);
\end{tikzpicture}
}; 

\node at (bbox.north west) (a) {\textbf{a)}};
\node at (a |- matrix.north) {\textbf{b)}};

\end{tikzpicture}%

\else
\includegraphics{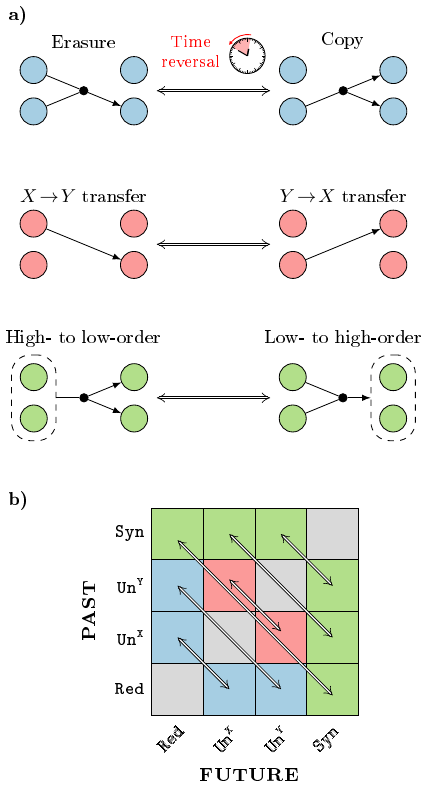}
\fi 

\caption{\textbf{Irreversible modes of information dynamics}. \textbf{a)} Examples of different irreversible modes of information dynamics and their time-reversed counterparts: irreversibility due to asymmetry between copy and erasure of information (top); irreversibility due to asymmetry of information transfer between the parts (middle); and high-order irreversibility due to asymmetric transfer of information between the whole and its parts (bottom).  
\textbf{b)} Effect of time reversal on the 16 information modes between two time series.} 
\label{fig:example}

\end{figure}
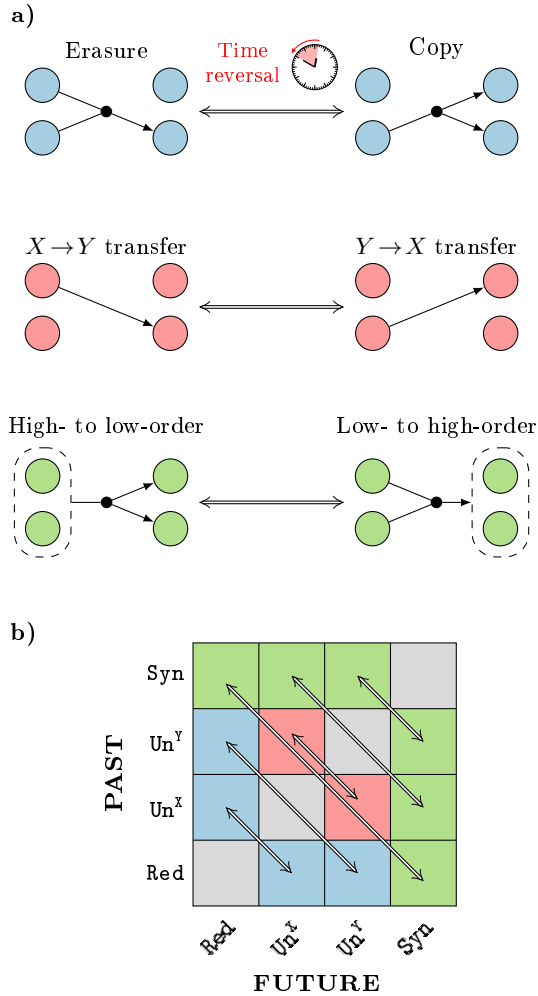

\vspace{0.5cm}

\paragraph{Irreversible modes of information dynamics.}
For simplicity, we begin  by considering a system described by the co-evolution of two interdependent time series $X_{t}$ and $Y_{t}$ under Markovian dynamics (see the Appendix for a more general treatment). 
Thanks to the Markov condition, the total amount of information shared between timepoints $t$ and $t'>t$ is captured by the mutual information between the system's state at those timepoints, i.e. $I(X_t,Y_t; X_{t'},Y_{t'})$~\cite{crutchfield2003regularities}. 

Crucially, this information can be decomposed into distinct modes of information dynamics depending on whether the information is held by only one of the parts, or each of them separately. More specifically, the framework of Integrated Information Decomposition (\phiid)~\cite{mediano2021towards}
proposes the following decomposition:
\begin{align}
    I(X_t Y_t; X_{t'} Y_{t'}) = \sum_{\substack{ \bm \alpha,\bm\beta \in A}} \PI{\bm\alpha}{\bm\beta},
\end{align}
where $A=\{\texttt{Red},\texttt{Un\textsuperscript{X}},\texttt{Un\textsuperscript{Y}},\texttt{Syn}\}$ is a collection of `information atoms.' Briefly, information that is found in either element of the system at a particular time is referred to as `redundant' (\texttt{Red}). Information that is only found in one element, but not the other, is termed `unique' ($\texttt{Un\textsuperscript{X}}$ and $\texttt{Un\textsuperscript{Y}}$ for $X_t$ and $Y_t$, respectively). And finally, information that is held by the system as a whole when the parts are considered jointly, but not by either of the parts in isolation, is termed `synergy' (\texttt{Syn}). In a time series context, the arrow denotes how this information evolves over time --- for example, \rts represents information that was redundant at time $t$ and synergistic at time $t'$, and similarly for all other atoms. For a brief introduction to $\Phi$ID and how to compute its terms, please refer to the Appendix.

What happens if one reverses the temporal ordering? To study this, let us consider time series $\tilde{X}_t$ and $\tilde{Y}_t$ with reversed temporal ordering, such that the transition $\big(\tilde{X}_t,\tilde{Y}_t\big) \to \big(\tilde{X}_{t'},\tilde{Y}_{t'}\big)$ is equivalent to 
$\big(X_{t'},Y_{t'}\big)\to\big(X_{t},Y_{t}\big)$. Then, a direct calculation shows that the total information that goes from $t$ to $t'$ for $\tilde{X}_t$ and $\tilde{Y}_t$ is the same as with $X_t$ and $Y_t$:
\begin{equation*}
    I\big(\tilde{X}_t \tilde{Y}_t; \tilde{X}_{t'} \tilde{Y}_{t'}\big) 
    = I\big(X_{t'} Y_{t'}; X_{t} Y_{t}\big) 
    = I\big(X_t Y_t; X_{t'} Y_{t'}\big) \,,
\end{equation*}
where the last step relies on the symmetry of Shannon's mutual information. However, a fundamental fact exploited by our formalism is that, despite the total information being equal, the information decomposition of $\tilde{X}_t$ and $\tilde{Y}_t$ is not the same as that of $X_t$ and $Y_t$.
A direct evaluation shows that time reversal of time series data makes information atoms transform into others as follows (Fig.~\ref{fig:example}):

\begin{itemize}
    \item[(i)] \textit{Erasure and copy of information}: \ytr exchanges with \rty, and \xtr exchanges with \rtx. Unique information becoming redundant means that the information is being duplicated, whereas the opposite means that information is being erased.
    \item[(ii)] \textit{Transfer reversal}: \xty exchanges with \ytx. When information that was uniquely held by one element, is then uniquely held by the other, that information has been transferred between the elements.
    \item[(iii)] \textit{High-order}: \sty exchanges with \yts, and \stx exchanges with \xts. Also, \rts exchanges with \str. These dynamics represent interactions between the system as a whole and its individual parts.
\end{itemize}
In contrast, the following atoms remain invariant to time reversal: \rtr, \xtx, \yty, and \sts. These modes represent information that is persistently held in the same way in the system, and hence are time-symmetric.

The three types of temporal asymmetries and corresponding transformations between information atoms outlined above provide a basis to distinguish between three qualitatively distinct modes of irreversibility. In effect, temporal irreversibility can occur either due to asymmetries between (i) erasure and copy, (ii) the direction of information transfer, or (iii) the influences between low- and high-order levels in a system. 
Moreover, \phiid provides natural metrics to capture the presence of these different types of temporal irreversibility:
\begin{align}
    \begin{rcases}
    i_\text{{\makebox[0pt][l]{ec}\hphantom{ec}}}^{x} =&  \rty - \ytr \\
    i_\text{{\makebox[0pt][l]{ec}\hphantom{ec}}}^{y} =&  \rtx - \xtr \quad \,\,
    \end{rcases}
    &\text{\small{ erasure-copy}} \nonumber\\
    \begin{rcases}
    i_\text{\makebox[0pt][l]{t}\hphantom{ec}} =& \xty - \ytx \quad \,\, \,\,
    \end{rcases}
    &\text{\small{ transfer}} \nonumber\\
    \begin{rcases}
    i_\text{hi}^{x} =& \stx - \xts \\
    i_\text{hi}^{y} =& \sty - \yts \\
    i_\text{hi}^{r} =& \rts - \str \quad
    \end{rcases}
    &\text{\small{ high-order}} \nonumber
\end{align}
In turn, we can build non-negative metrics of the degree of irreversibility of each type as follows:

\begin{equation}
\begin{aligned}
    I_\text{\makebox[0pt][l]{ec}\hphantom{ec}} =& ~|i_\text{ec}^{x}| + |i_\text{ec}^{y}|, \\
    I_\text{\makebox[0pt][l]{t}\hphantom{ec}} =& ~|i_\text{t}| , \\
    I_\text{\makebox[0pt][l]{hi}\hphantom{ec}} =& ~|i_\text{hi}^{x}| + |i_\text{hi}^{y}| + |i_\text{hi}^{r}| .
\end{aligned}
\label{eq:metrics_modes}
\end{equation}

In summary, a system has irreversible dynamics if $\max\{I_\text{ec}, I_\text{t}, I_\text{hi}\} > 0$. 
Importantly, more than one mode of irreversibility may be present, and different systems may exhibit the same overall level of irreversibility but due to the prevalence of different modes. See the Appendix for detailed instructions on how to compute the \phiid information atoms and these irreversibility metrics.

\vspace{0.4cm}
\paragraph{Decomposing heuristic markers of irreversibility.}
To demonstrate the capabilities of our framework, we next show how it can be used to derive an exact information-theoretic decomposition of INSIDEOUT, a heuristic marker for temporal irreversibility that has proven to be effective at characterising different brain states and pathologies~\cite{deco2022insideout}. Briefly, the proposed marker is applied to pairs of timeseries, and it involves calculating the cross-correlations between a pair of timeseries (e.g. the activity of two brain regions), and comparing its value to the one obtained with the time-reversed timeseries. Under the assumption (adopted in Ref.~\cite{deco2022insideout}) that the corresponding variables are normally distributed, one can re-write the INSIDEOUT metric of irreversibility as:
\begin{equation}\label{eq:def_insideout}
    \text{INSIDEOUT} = | I(X_t;Y_{t'}) - I(Y_t;X_{t'}) | ^2~.
\end{equation}
A direct calculation (see Appendix) shows that this proxy can be expressed in terms of elementary modes of irreversibility, as derived above, as
\begin{align}
\text{INSIDEOUT} = |i_\text{ec}^{x} + i_\text{ec}^{y} + i_\text{t} |^2.
\label{eq:insideout_phiid}
\end{align}

This result reveals that $\text{INSIDEOUT}$ is a mixture of two distinct sources of irreversibility: transfer reversal and erasure-copy. This is important, as if these types of irreversibility were of opposite signs they could cancel each other --- thus, a system with information dynamics that hold different modes of irreversibility may falsely appear reversible under this proxy. Such scenarios can be avoided by capturing the contribution of each irreversibility mode separately using, e.g. the formulas in Eq.~\eqref{eq:metrics_modes}. 

Our framework also reveals that data-generating assumptions may limit the types of irreversibility that one can observe. For example, several variants of information decomposition are equivalent to the Minimum Mutual Information (MMI) definition of information redundancy if one assumes that the data follows a multivariate Gaussian distribution~\cite{barrett2015exploration}. Under such conditions, then it can be shown (proof in the Appendix) that $i_\text{ec}^{x} = i_\text{ec}^{y} = 0$, showing that Gaussian systems do not have erasure-copy irreversibility. Therefore, under the normality assumption, INSIDEOUT reduces to
\begin{equation}
\text{INSIDEOUT} = | \xty - \ytx |^2 = I_t^2 ~ ,
\label{eq:infotransfer}
\end{equation}
showing that, under the assumption of Gaussian data, INSIDEOUT captures exclusively the asymmetry in information transfer between elements. 
This finding also provides a formal explanation for the empirical correlations between temporal irreversibility as quantified by INSIDEOUT and transfer entropy applied to functional brain activity data (reported e.g. in Refs.~\cite{kringelbach2023toward, deco2022insideout}). More details of this relationship are provided in the Appendix.

\begin{figure*} 
\centering

\if1\compiletikz
  \tikzsetnextfilename{ResultsFigure}%
  \filemodCmp{tikz/ResultsFigure.tex}{tikz/external/ResultsFigure.pdf}%
    {\tikzset{external/remake next}}{}%
  \input{tikz/ResultsFigure.tex}%

\else
\includegraphics{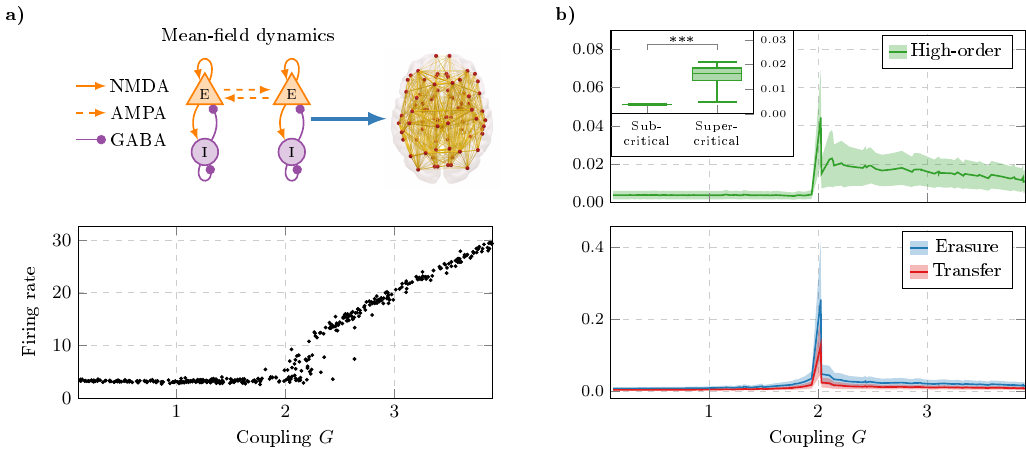}
\fi 

\caption{\textbf{Distinct modes of temporal irreversibility in simulated brain dynamics.} \textbf{a)} A biophysical model of excitatory (E) and inhibitory (I) neural masses coupled according to the structural connectivity of the human brain is used to simulate the neural dynamics of different brain regions. The model displays a transition from a low- to a high- (uncontrolled) firing rate regime as the global coupling parameter $G$ increases, with a sharp transition at approximately $G=2$. 
\textbf{b)} Strength of the different modes of irreversibility as function of the global coupling parameter. Shaded area represents the inter-quantile range. All models exhibit a peak at the phase transition, and high-order irreversibility dominates in the high-activity regime.
}
\label{fig_Modes}
\end{figure*}

\vspace{0.5cm}
\paragraph{High-order irreversibility.}
While it is relatively straightforward to reason about temporal irreversibility due to asymmetries in information transfer or copy and erasure processes (captured by $I_\text{ec}$, $I_\text{t}$, and the INSIDEOUT metric), our framework also reveals a new, previously unreported source of irreversibility pertaining to synergistic information. 
This mode of reversibility (captured by $I_\text{hi}$) refers to the flow of information between the parts and the whole system: specifically, it compares the amount of information held by the parts at time $t$ 
and becomes encoded in the whole at time $t'$, versus information that was encoded in the whole at time $t$ and is encoded in the parts at time $t'$. 
We therefore refer to this as a \emph{high-order asymmetry} because it bridges levels of the part-whole hierarchy, unlike the `horizontal' asymmetries captured by $I_\text{ec}$ and $I_\text{t}$ that are solely related to relationships between parts. 

Hence, our formalism highlights and enables us to capture a previously undescribed phenomenon: a system that appears reversible when only part-part relationships are taken into account may still be overall irreversible due to high-order asymmetries in the dynamics between parts and whole. Moreover, as we show below, far from being a mere theoretical possibility, high-order irreversibility can play an important role in a system's dynamics.

\vspace{0.5cm}
\paragraph{Simulated brain dynamics.}
We applied these ideas to the analysis of neural dynamics generated by a well-known computational whole-brain model. The model is a biophysically plausible dynamic mean field (DMF) model, consisting of neural masses representing populations of excitatory and inhibitory neurons coupled according to the network structure of the human brain~\cite{deco2018lsd, herzog2022dmf}. It has been shown that the model exhibits an abrupt transition for a critical value of a control parameter $G$, which represents the global coupling strength between brain regions~\cite{deco2018lsd, herzog2022dmf}. We studied the irreversibility of the pairwise brain dynamics obtained at various levels of $G$ above and below the critical value (Fig.~\ref{fig_Modes}): for each pair of regional timeseries, we calculated the value of each of the irreversibility modes, as well as INSIDEOUT for comparison.

Results show that while while the system exhibits peaks of irreversibility at the critical threshold in all three modes, only the high-order asymmetry $I_\text{hi}$ effectively differentiates between the subcritical and supercritical regimes. In particular, there is significantly greater high-order irreversibility in the supercritical regime than in the subcritical regime (subcritical: mean $= 0.0038$, SD $= 1.27 \times 10^{-4}$; supercritical: mean $= 0.0162$, SD $= 0.0029$; $t(95) = 30.41$;  Hedge's $g = 6.31$ (95$\%$ CI: $5.26$ to $7.62$); $p < 0.001$; Fig.~\subref{fig_Modes}{b}, inset). 
This difference is significantly greater than the subcritical-vs-supercritical difference observed for either kind of lower-order irreversibility, $I_\text{ec}$ and $I_\text{i}$ (both $p<0.001$, $Z = 6.41$ for difference between effect sizes). The  difference in high-order irreversibility between subcritical and supercritical regimes is also significantly greater than the corresponding difference obtained in a null model, in which the temporal order of activity is randomly reshuffled (removing all temporal correlations) while preserving the covariance:  $p<0.001$, $Z = 8.99$ for difference between effect sizes. 
Additionally, our results also confirm that Eq.~\eqref{eq:infotransfer} and INSIDEOUT provide numerically identical results (Spearman $\rho = 1.0$, $p < 0.001$).

\vspace{0.5cm}
\paragraph{Discussion.}
The presented framework reveals a more nuanced understanding of the arrow of time in multivariate systems, where --- rather than a monolithic entity --- it is conceived as a composite of qualitatively distinct modes of information dynamics. 
For the case of two time series, our results identified three types of dynamics that can generate temporal irreversibility, which can coexist in different degrees. 
Crucially, this means that different systems may exhibit the same degree of overall irreversibility, but this may be driven by qualitatively different types of dynamics --- a possibility that was not appreciated until now. 
 
In other words, assessing temporal irreversibility via a single scalar can obscure differences between systems, which our decomposition brings to light. Perhaps most surprisingly, our framework revealed the existence of `high-order' modes of temporal irreversibility, which arise from asymmetric relationships between the whole system and its parts. 
Importantly, it is possible for a system to have high-order irreversibility while appearing reversible from its part-to-part relationships. 

As a proof of concept, we found evidence that high-order irreversibility dominates in the supercritical regime of a biophysical model of human brain activity. This result demonstrates that high-order irreversibility is not only a theoretical possibility, but a genuine phenomenon relevant to widely-studied systems.

The distinction between the different types of dynamics that give rise to irreversibility provides a common language to analyse and compare existent measures of irreversibility. As an illustration of this,  we showed that the INSIDEOUT marker captures only asymmetries in the information transfer between parts of the system. 
Moreover, having irreversible dynamics was shown to be a sufficient condition to having a non-zero entropy production. By leveraging the multivariate structure of the system, this yields computationally efficient ways to evaluate the irreversible of a system that are of great attractive for practical data analysis.

One important feature that makes our framework complementary to previous approaches based on entropy production (e.g. Refs.~\cite{ito2020unified,lynn2022decomposing, seif2021machine}) is that our framework is based on decomposing the information transferred between the past and the future states of a multivariate system. 
Formalising the relationship between these two types of approaches is not straightforward, and remains a promising avenue for future research. 
The approach presented here is particularly well-suited to practical data analysis, as it does not rely on assumptions (such as multipartite dynamics) which often do not hold in empirical time series data. Overall, we have demonstrated theoretically and empirically that the same levels of temporal irreversibility can arise from by qualitatively different types of dynamics, and we provide practical means of disentangling them through information decomposition.

\paragraph{Acknowledgements.}
A.L. is supported by the Molson Neuro-Engineering Fellowship of the Montreal Neurological Institute and UNIQUE Neuro-AI Excellence Award. F.R. was supported by the Fellowship Programme of the Institute of Cultural and Creative Industries of the University of Kent, and the DIEP visitors programme at the University of Amsterdam.  

\putbib
\end{bibunit}

\appendix
\section{APPENDIX}

\begin{bibunit}

\subsection*{Brief introduction to \phiid}

Integrated Information Decomposition (\phiid;~\cite{mediano2021towards}) is a generalisation of Partial Information Decomposition (PID;~\cite{williams2010nonnegative}), a technique to decompose the information that multiple \textit{source} variable have about a single \textit{target} variable. \phiid generalises PID by extending the formalism to systems with multiple sources \textit{and} multiple targets. 
This makes \phiid particularly well-suited for the study of multivariate stochastic processes, where the state of each process at time $t$ can be seen as the sources of information and the states at $t'>t$ as targets. In such scenarios, \phiid describes how information is carried by the variables in the system over time. 

For the case of two sources of information $X,Y$ and one target $Z$, PID divides their information into four ``information atoms:'' redundancy ($\texttt{Red}$), synergy ($\texttt{Syn}$), and unique of the first and second sources ($\texttt{Un\textsuperscript{X}}$ and $\texttt{Un\textsuperscript{Y}}$). The \phiid of two timeseries $(\dots,X_t,X_{t+1},\dots)$ and $(\dots,Y_t,Y_{t+1},\dots)$ between timepoints $t$ and $t'$ considers two sources of information $X_t,Y_t$ and two targets $X_{t'},Y_{t'}$, and their information is divided into 16 atoms that comprise all possible ways in which the 4 atoms of traditional PID can evolve in time. For example, \phiid's \xty atom is the information transferred from $X_{t}$ to $Y_{t'}$ --- i.e. information that was uniquely present  in $X$ at time $t$, and is present only in $Y$ at time $t'$. Similarly, \rty is a case of information \textit{erasure}: it was present in both $X$ and $Y$ (i.e., redundant) at time $t$, and is present only $Y$ (and hence not anymore in $X$) at time $t'$. As another example, \rtr is information that is persistently redundant: it was present redundantly in both $X$ and $Y$ at time $t$, and stays redundant at time $t'$. Similarly, \sts was carried synergistically by $X$ and $Y$ at time $t$, and is persistently synergistic at time $t'$. More details about \phiid can be found in Ref.~\cite{mediano2021towards}.

Given a time series dataset from a bivariate stochastic process, it is possible to compute all atoms in its \phiid decomposition given a suitable form of the joint probability density function $p(X_t, Y_t, X_{t'}, Y_{t'})$. In this work, we take the simple yet empirically useful assumption that this probability density is a multivariate Gaussian distribution (which is an accurate description of functional brain activity data~\cite{nozari2020brain}).

Under these assumptions, the mutual information between two random variables $U$ and $V$ can be computed from their joint covariance matrix as
\begin{align}
    I(U; V) = \frac{1}{2} \log \frac{|\Sigma_U| |\Sigma_{V}|}{|\Sigma_{UV}|} ~ ,
    \label{eq:mi}
\end{align}
\noindent where $\Sigma_U$ (resp. $\Sigma_V$) is the covariance matrix of $U$ (resp. $V$), and $\Sigma_{UV}$ is the joint covariance matrix of $U$ and $V$. With this expression, we can compute the mutual information between any pair of (sets of) variables in the data. As an example, the time-delayed mutual information of the joint stochastic process is given by
\begin{align}
    I(X_t, Y_t; X_{t'}, Y_{t'}) = \frac{1}{2} \log \frac{|\Sigma_{X_t Y_t}| |\Sigma_{X_{t'}Y_{t'}}|}{|\Sigma_{X_tY_tX_{t'}Y_{t'}}|} ~ .
    \label{eq:tdmi}
\end{align}
Equipped with this expression for mutual information, we can now tackle the problem of computing the atoms of PID and \phiid. Under reasonable assumptions~\cite{barrett2015exploration,venkatesh2022partial}, the PID of Gaussian variables reduces to the so-called Minimum Mutual Information (MMI) approach, under which redundancy is calculated as
\begin{equation}
    \texttt{Red}(X,Y; Z) = \min\{I(X;Z),I(Y;Z)\} ~ 
    \label{eq:mmi}
\end{equation}
\noindent for any target variable $Z$. Similarly, in the case of \phiid, this redundancy measure can be extended to yield the double-redundancy atom,
\begin{multline}
    \rtr = \min\{I(X_t; X_{t'}), I(X_t; Y_{t'}), \\ I(Y_t; X_{t'}), I(Y_t; Y_{t'})\} ~ .
    \label{eq:mmi_doublered}
\end{multline}
Finally, after calculating \rtr with Eq.~\eqref{eq:mmi_doublered}, PID redundancy with Eq.~\eqref{eq:mmi}, and mutual information with Eq.~\eqref{eq:mi} for all relevant system subsets, the other 15 information atoms can be calculated as the solution of a simple linear system of equations, as described in Ref.~\cite{mediano2021towards}. The reader is referred to the original manuscript and the supplementary software provided with Ref.~\cite{luppi2022synergistic} for a description and open-source code to compute MMI \phiid from time series data.

\subsection*{Irreversible information dynamics\\in multivariate systems}

In the main text, we have introduced a distinction between three modes of irreversibility for a bivariate stochastic process.\footnote{Specifically, the important condition is that the joint process must be partitioned into two interdependent processes $X_t$ and $Y_t$, although these need not be scalar-valued.} Here we describe the more general scenario of a multivariate system with $n \geq 2$ parts, and show that an analogous division applies.\footnote{The presented formalism can also be readily extended to non-Markovian dynamics, which will be explored in future work.}

Consider the set of all integrated information atoms for $n$ time series, $\mathcal{S} = \{\alpha\to\beta ~ | ~ \alpha,\beta \in \mathcal{A}\}$, where $\mathcal{A}$ is the set of antichains of $\{1, ..., n\}$, following prior work~\cite{mediano2021towards,williams2010nonnegative}. In this notation, the atom \rtr is written as $\{1\}\{2\}\to\{1\}\{2\}$, the atom \stx as $\{1,2\}\to\{1\}$, and similarly for all others. The set of atoms that remain invariant under time reversal is $\mathcal{R} = \{\alpha\to\beta ~ | ~ \beta = \alpha\} \subset \mathcal{S}$, and removing these we are left with the set $\mathcal{I} = \mathcal{S} \setminus \mathcal{R}$ of atoms that in general do change with time reversal. We are now in a position to formulate our general irreversibility decomposition.

First, we consider atoms that constitute erasure or copy of information. For this, we note that if $\alpha\subset\beta$ then the information in $\alpha$ stays in those sources and additionally it gets copied to the sources in $\beta$ that were not in $\alpha$ --- which constitutes a case of information duplication. Analogously, if $\beta\subset\alpha$ then the information in sources in $\alpha$ that are not in $\beta$ disappear, which constitutes information erasure. Hence, the set of all atoms related to the erasure-copy class is given by
\begin{equation}
    \mathcal{I}_\text{ec} = \{ \alpha\to\beta | \alpha\subset \beta \text{ or } \beta\subset\alpha\}~.
\end{equation}

Next, let's consider atoms that constitute cases of information transfer. For this, let's recall the definition of precedence in PID: $\alpha$ is said to precede $\beta$ in the PID lattice if all sources in $\alpha$ are contained in at least one source in $\beta$~\cite{williams2010nonnegative}. Conversely, $\alpha$ does not precede $\beta$ if at least one source in $\alpha$ is not contained in any source of $\beta$. Building on this, we say that $\alpha\to\beta$ refers to transfer if neither $\alpha$ or $\beta$ precede the other --- hence, transfer takes place between sources that are not contained in each other. Then, the set of all atoms associated with information transfer are
\begin{equation}
    \mathcal{I}_\text{t} = \{ \alpha\to\beta | \alpha\npreceq \beta \text{ and } \beta\npreceq \alpha\}~.
\end{equation}

Finally, let's characterise the remaining atoms. If an atom is not in $\mathcal{I}_\text{ec}$ or $\mathcal{I}_\text{t}$, then one element precedes the other but the sources of the preceding atom are different from those in the other (so it is not erasure or copying). This can only happen if there is at least one synergistic source in the largest, which is broken down into redundant or unique parts in the smaller one (e.g. $\{1,2\}\to\{1\}\{2\}$). This interaction between synergies involving groups of variables and parts of them can be thought of as information flowing between orders, or between different scales, of the system --- either from macro-to-micro or vice versa. We call these ``cross-order'' modes of information, and can be defined as follows:
\begin{align}
    \mathcal{I}_\text{c} = \big\{ \alpha\to\beta 
    \big| 
    (\alpha\preceq \beta 
    \text{ and } 
    \alpha\not\subset \beta) 
    \text{ or } 
    (\beta\preceq \alpha
    \text{ and }
    \alpha\not\subset \beta) 
    \big\}. \nonumber
\end{align}
It is direct to verify that $\mathcal{I}_\text{ec}$, $\mathcal{I}_\text{t}$, and $\mathcal{I}_\text{c}$ form a partition of $\mathcal{I}$, i.e. $\mathcal{I}_\text{ec} \cap \mathcal{I}_\text{t} = \mathcal{I}_\text{ec} \cap \mathcal{I}_\text{c} = \mathcal{I}_\text{t} \cap \mathcal{I}_\text{c} = \varnothing$, and $\mathcal{I}_\text{ec} \cup \mathcal{I}_\text{t} \cup \mathcal{I}_\text{c} = \mathcal{I}$. Therefore, these three sets constitute a proper decomposition of the irreversible information flow in the system's dynamics.

So far we have distinguished three classes of irreversible information dynamics --- related to copy-erasure, transfer, and cross-order phenomena. 
A useful further distinction is whether these atoms are low- or high-order. For this, we take inspiration from prior work~\cite{rosas2020reconciling,ehrlich2023a} and define the order of a \phiid atom as 
\begin{equation}
    s(\alpha\to\beta) = \max_{\gamma\in\{\alpha,\beta\}} \Big\{ \max_{g\in\gamma} |g| \Big\}.
\end{equation}
In simple words, $s(\alpha\to\beta)$ is the cardinality of its largest (synergistic) source. 
It can be shown that atoms in $\mathcal{I}_\text{c}$ are always of order 2 or more, and hence are always of high-order. In contrast, atoms in $\mathcal{I}_\text{ec}$ and $\mathcal{I}_\text{t}$ can be of low or high order. This leads to a decomposition into five classes of irreversible atoms: 
\begin{itemize}
    \item low- and high-order transfer, 
    \item low- and high-order erasure-copy, and
    \item cross-order phenomena.
\end{itemize}
Note that for the particular case of $n=2$ time series all transfer and erasure-copy atoms are of low-order, and the only high-order atoms are of the cross-order type.

\subsection{Irreversible information dynamics imply non-zero entropy production}

Here we provide a formal proof that having irreversible dynamics (i.e. that $\max\{I_\text{ec},I_\text{t},I_\text{hi}\}>0$) is a sufficient condition for a system to have non-zero entropy production. To prove this via a contrapositive argument, we start by assuming that a system has zero entropy production and we show that this implies that it has no irreversible dynamics. For simplicity of the proof, we focus on systems with discrete-time stationary Markovian dynamics, for which the entropy production $\Sigma$ can be calculated as~\cite{igarashi2022entropy}
\begin{align}
    \Sigma 
    &= D\big(p(X',Y'|X,Y)||q(X',Y'|X,Y)\big) \\
    &= \mathbb{E}\left\{ \log \frac{p(X,Y|X',Y')}
    {p(X',X'|X,X)} \right\}~,
\end{align}
where $D(p||q)$ is the Kullback-Leibler divergence,  $q(x,y|x',y')=p(x',y'|x,y)$ is the distribution of the backward process, and $X=X_t,X'=X_{t+1},Y=Y_t$ and $Y'=Y_{t+1}$ is a shorthand notation. 

If $\Sigma=0$, then $p = q$, which implies that $p(x,y|x',y') = p(x',y'|x,y)$ for all $x,y,x'$ and $y'$. This, combined with the assumption of stationarity, implies that any statistic that one could compute over $(X,Y,X',Y')$ will yield the same value if one swaps $(X,Y)$ for $(X',Y')$. In particular, each of the \phiid atoms have the same value after time reversal. This implies, in turn, that $\max\{I_\text{ec},I_\text{t},I_\text{hi}\}=0$, concluding the proof.

This result confirms that our framework establishes practical ways to determine irreversibility without the need of calculating the entropy production. In essence, our framework leverages the multivariate structure of a system to enable computational shortcuts to evaluate the system's irreversibility. For example, if one needs to confirm whether a system is irreversible or not, a simple calculation such as the one required to estimate INSIDEOUT (see Eq.~\eqref{eq:def_insideout}) can suffice to provide sufficient proof of irreversibility. This opens multiple new avenues for practical analyses of irreversibility in real-world applications.

\subsection{Derivation of INSIDEOUT decomposition}

The INSIDEOUT measure quantifies temporal irreversibility by considering pairs of timeseries --- specifically, through the asymmetry of the time-delayed mutual information between forward and time-reversed versions of the timeseries~\cite{deco2022insideout}.
The starting point for INSIDEOUT is the time-lagged mutual information between Gaussian timeseries X and Y at timepoints $t$ and $t'=t+\tau$, which we denote by $\text{FS}_\text{f}$ (for shifted FC) following the original notation~\cite{deco2022insideout}, and is given by
\begin{align}
\text{FS}_\text{f} = I(X_t; Y_{t'}) -\frac{1}{2} \log\big(1-\langle X_{t}, Y_{t'}\rangle^2\big)\,, \nonumber
\end{align}
where $\langle X, Y\rangle$ denotes the Pearson correlation between $X$ and $Y$. Analogously, the time-lagged mutual information between the time-reversed timeseries (indicated as $\tilde{X}_t$ and $\tilde{Y}_t$, respectively) can be calculated as
\begin{align*}
\text{FS}_\text{r} = I(\tilde{X}_t, \tilde{Y}_{t'}) = -\frac{1}{2} \log\big(1-\langle \tilde{X}_{t}, \tilde{Y}_{t'}\rangle^2\big) ~ .
\end{align*}
In turn, the INSIDEOUT marker of temporal irreversibility is defined as
%
\begin{align}
\text{INSIDEOUT} = \big(\text{FS}_\text{f} - \text{FS}_\text{r}\big)^2\,. \nonumber
\end{align}
In other words, this marker of temporal irreversibility captures the difference between the time-lagged mutual information between these X and Y timeseries in the forward and reversed direction~\cite{deco2022insideout}.

Importantly, computing the time-delayed mutual information as a function of time-shifted correlation implies assuming Gaussianity. Under this assumption, it can be shown that $\text{FS}_\text{f} = I(X_t;Y_{t'})$ and 
$\text{FS}_\text{r} = I(Y_t;X_{t'})$.

Having expressed $\text{FS}_\text{f}$, $\text{FS}_\text{r}$, and INSIDEOUT in terms of mutual information, we can now decompose them into their constituent information atoms using \phiid. Specifically, $\text{FS}_\text{f}$ and $\text{FS}_\text{r}$ can be written as follows:
\begin{align*}
\text{FS}_\text{f} &= \rtr + \rty + \xtr + \xty ~, \\
\text{FS}_\text{r} &= \rtr + \rtx + \ytr + \yty ~ .
\end{align*}
This is because \rtr is information that is persistently redundant: it was present in both $x$ and $y$ at $t$, and is also present in both $x$ and $y$ at $t'$; therefore, it satisfies the condition of being present both in $x$ at $t$, and the condition of being present in $Y_{t'}$, which means that it is part of $\text{FS}_\text{f}$; 
\rty is information that was redundant, and is now unique to y. It was present in both x and y at t, and is present in y at t'; therefore, it satisfies our conditions for being part of $\text{FS}_\text{f}$; 
\xtr is information that was unique to x, and is now redundant. It was present in x at t, and is present in both x and y at t'; therefore, it satisfies our conditions for being part of $\text{FS}_\text{f}$;
and finally, \xty is the pure transfer from $X_{t}$ to $Y_{t'}$, i.e. information that was present only in x at t, and is present only in y at t'; therefore, it satisfies our conditions for being part of $\text{FS}_\text{f}$.

Finally, using these expressions, and given that
$\text{INSIDEOUT}= (\text{FS}_\text{f} - \text{FS}_\text{r})^2$, a direct calculation shows that

\begin{align}
\text{INSIDEOUT} =& \big(  \rty + \xtr + \xty   \nonumber\\
&- \ytr - \rtx - \ytx \big)^2 ~ .\nonumber
\end{align}

Recalling the definition of irreversibility metrics from the main text, by simply matching pairs of atoms in the equation above it is now straightforward to re-write INSIDEOUT as
\begin{align*}
\text{INSIDEOUT} = |i_\text{ec}^{x} + i_\text{ec}^{y} + i_\text{t} |^2 ~ ,
\end{align*}
\noindent recovering the result of Eq.~\eqref{eq:insideout_phiid}.

\subsection{Information decomposition of NDTE}

In Ref.~\cite{deco2021revisiting}, Deco and colleagues define the Normalised Directed Transfer Entropy, $\text{NDTE}(X\to Y)_t^{t'}$, as a measure of the directed information flow between two brain regions $X$ and $Y$ from time $t$ to time $t'$. In this section we leverage the \phiid framework to explain the empirical correlation between NDTE and INSIDEOUT~\cite{deco2022insideout}.

Instead of NDTE per se, for simplicity we consider its unnormalised version, which is simply the standard transfer entropy from $X$ to $Y$, given by
\begin{equation}
    \text{TE}(X\to Y)_t^{t'}:=I(Y_{t'};X_{t}|Y_{t})~,
    \label{eq:te}
\end{equation}
\noindent which quantifies the information that is present in $Y_{t'}$ and was already present in $X_t$, but not in $Y_t$~\cite{schreiber2000measuring}. Following previous work by Williams and Beer \cite{williams2010nonnegative}, Mediano and colleagues \cite{mediano2021towards} showed that TE can be decomposed in terms of information atoms as 
\begin{align}
    \text{TE}(X\to Y)_t^{t'} =& ~  \xty + \sty \\
    &+ \str + \xtr \nonumber ~ .
\end{align}
Taking the difference in TE between both directions of information flow ($X\to Y$ and $Y\to X$) we arrive at the following expression:
\begin{align*}
    \text{TE}(X\!\to \!Y)_t^{t'} \!-\! \text{TE}(Y\!\to \!X)_t^{t'} \!&= i_t\\ &+  (\sty \!-\! \stx)\\ &+  (\xtr \!-\! \ytr) .
\end{align*}

This decomposition lays bare the origin of the empirical relationship previously observed between (ND)TE and INSIDEOUT. Namely, both metrics include $i_t$, the transfer component of temporal irreversibility. However, NDTE is an imperfect proxy for INSIDEOUT, because its two other terms consider spatial asymmetry (between $X$ and $Y$) instead of temporal asymmetry (between past and future). Thus, through the lens of information decomposition we can express both INSIDEOUT and TE in a common language, revealing both the similarities and the differences between them, and clarifying which aspects of the system they discern.

\subsection{Consequences of the MMI decomposition for INSIDEOUT}

As described in the first section of this Appendix and in Eq.~\eqref{eq:mmi}, for a multivariate Gaussian distribution the Minimum Mutual Information (MMI) redundancy function is given by
\begin{equation*}
    \text{Red}(X,Y;Z) = \min\{I(X;Z),I(Y;Z)\} ~ .
\end{equation*}
Let us introduce the following shorthand notation:
\begin{alignat*}{3}
    &A_x &:= I(X_t; X_{t'}), \\
    &A_y &:= I(Y_t; X_{t'}),\\
    &R_{xy} &:= I(X_{t};Y_{t'}),\\ 
    &R_{yx} &:= I(Y_t; X_{t'}).
\end{alignat*}
Let us assume that the autocorrelations are stronger than the cross-correlations, i.e. that
\begin{equation*}
    \min\{A_x,A_y\} \geq \max\{ R_{xy},R_{yx} \} ~ ,
\end{equation*}
\noindent which is an uncontroversial assumption in empirical and simulated fMRI data.

Then, the following redundancies can be calculated as follows:
\begin{alignat*}{3}
    &\text{Red}(X_t, Y_t; X_{t'}) &= R_{yx} \\
    &\text{Red}(X_t, Y_t; Y_{t'}) &= R_{xy} \\
    &\text{Red}(X_{t'}, Y_{t'}; X_t) &= R_{xy} \\
    &\text{Red}(X_{t'}, Y_{t'}; Y_t) &= R_{yx}
\end{alignat*}
Therefore, there are only two different redundancy values.

And now some \phiid atoms with the minimum mutual information (MMI) redundancy function:
\begin{align*}
    \rtr &= \min\{R_{xy}, R_{yx}\} \\
    \rtx &= \text{Red}(X_t, Y_t; X_{t'}) - \rtr \nonumber\\
    &= R_{xy} - \rtr \\
    \rty &= \text{Red}(X_t, Y_t; Y_{t'}) - \rtr \nonumber\\
    &= R_{yx} - \rtr \\
    \xtr &= \text{Red}(X_{t'}, Y_{t'}; X_t) - \rtr \nonumber\\
    &= R_{xy} - \rtr \\
    \ytr &= \text{Red}(Y_{t'}, Y_{t'}; Y_t) - \rtr \nonumber\\
    &= R_{yx} - \rtr.
\end{align*}
Note that $\rtx = \xtr$ and $\rty = \ytr$, and hence with the MMI decomposition the copy-erasure mode of irreversibility is always zero, i.e. $i_\text{ec}^{x}=i_\text{ec}^{y}=0$. Naturally, this in general is not the case with other redundancy functions different from Eq.~\eqref{eq:mmi}, which will be especially relevant for systems with discrete states or nonlinear dynamics.

\subsection*{Dynamic mean-field model}

We simulated biophysically realistic time-series of brain activity using the well-known Dynamic Mean Field (DMF) model. The reader is referred to Refs.~\cite{deco2018lsd, herzog2022dmf} for details of the DMF model and its implementation. Due to its multi-platform compatibility, low memory usage, and high speed, we used the recently developed \texttt{FastDMF} library~\cite{herzog2022dmf}. Code is freely available online at  \url{https://www.gitlab.com/concog/fastdmf}. 

Briefly, the model simulates each brain region as a macroscopic neural field comprising mutually coupled excitatory and inhibitory populations (80\% excitatory and 20\% inhibitory), providing a neurobiologically plausible account of regional neuronal firing rate. Regions are connected according to empirical anatomical connectivity reconstructed from in-vivo diffusion MRI tractography. We  used diffusion  MRI  (dMRI)  data  from  the  $100$  unrelated  subjects (54 females and 46 males, mean age $= 29.1 \pm 3.7$ years)  of  the  HCP  $900$  subjects  data  release~\cite{vanessen2013neuroimg, glasser2013neuroimg}. All  HCP  scanning  protocols  were  approved  by  the  local  Institutional  Review  Board  at  Washington  University  in  St.  Louis. We refer the reader to Ref.~\cite{luppi2021networkneurosci} for detailed description of the reconstruction and tractography of each subject's dMRI data. We obtained a consensus structural connectivity matrix $A$ for use in the DMF model by following the procedure described by Ref.~\cite{Wang2019sciadv}, as follows: for each pair of regions $i$ and $j$, if more than half of subjects had non-zero connection $i$ and $j$, $A_{ij}$ was set to the average across all subjects with non-zero connections between $i$ and $j$. Otherwise, $A_{ij}$ was set to zero. 

The DMF model has one free parameter, known as ``global coupling'' and denoted by $G$, which modulates the overall strength of the connections between brain regions. We simulated brain activity at each value of $G$ between 0.1 and 4, using increments of 0.1. A Balloon-Windkessel haemodynamic model~\cite{friston2003dcm} was then used to turn simulated regional neuronal activity into simulated regional BOLD signal, obtaining 600 time-points for each region. These simulated regional BOLD signals were bandpass filtered in the same range as is typical for empirical data ($0.008-0.09$ Hz). Finally, we computed information decomposition and the INSIDEOUT marker between each pair of time-series, for each value of $G$, and aggregated the results by averaging across all pairs of regions.

\putbib
\end{bibunit}

\end{document}